\documentclass[12pt,preprint]{aastex}

\usepackage{epsfig,amsmath,subfigure}

\shortauthors{Bower et al.}
\shorttitle{3C 286}

\begin{document}

\title{A Search for Radio Transients in VLA Archival Images of the 3C 286 Field}

\author{Geoffrey C. Bower\altaffilmark{1} and Destry Saul\altaffilmark{2}}
\altaffiltext{1}{University of California, Berkeley, Radio Astronomy Laboratory and Department
of Astronomy, 601 Campbell Hall \#3411, Berkeley, CA 94720, USA; gbower@astro.berkeley.edu }
\altaffiltext{2}{Columbia University, Department of Physics, 538 West 120th Street, 1333 Pupin, 
MC 5255, New York, NY 10027; destry@astro.columbia.edu}

\begin{abstract}
We present a search for radio transients in the field of the bright radio source 3C 286
using archival observations from the Very Large Array.  These observations span 23 years
and include 1852 epochs at 1.4 GHz in the C and D configurations.  We find no
transients in the field.  The sensitivity of the observations is limited by dynamic range
effects in the images.  At large flux densities ($> 0.2$ Jy), single epoch observations provide a 
strong limit on the transient surface density.  At flux densities near the dynamic range threshold, we use the requirement
that transient sources must appear in consecutive epochs to be confirmed as real.  This
sets the sensitivity at low flux densities to transient durations of $\tau \sim 1$ day,
while  $\tau > 1$ minute for high flux densities.  At 70 mJy, we find a 1-$\sigma$ limit on the
surface density $\Sigma < 3 \times 10^{-3}$ deg$^{-2}$.  At 3 Jy, we find a 1-$\sigma$ limit 
$\Sigma < 9 \times 10^{-4}$ deg$^{-2}$.  A future systematic search of the VLA archives can provide
one to two orders of magnitude more sensitivity to radio transients.
\end{abstract}
\keywords{radio continuum:  general --- radio continuum:  stars ---
radio continuum:  galaxies --- surveys}

\section{Introduction}

Radio transient (RT) sources probe the high energy population of the Universe. 
Known hosts to transient radio emission include neutron stars, black holes, 
supernovae, gamma-ray bursts, and highly magnetized stars and planets.
Most of what is known about RTs has been learned from follow-up of 
events discovered at high-energy or optical wavelengths or through
serendipitous discovery \citep[e.g.,][]{2009A&A...499L..17B}.  
Systematic searches for RTs have been conducted but vast parameter space remains unexplored.

Blind searches for RTs are an important scientific goal
for major new radio telescope facilities such as the Allen Telescope 
Array \citep{2010ApJ...719...45C,2010ApJ...725.1792B}, LOFAR \citep{2009ASPC..407..318H}, 
ASKAP \citep{2010PASA...27..272M}, and the Long Wavelength Array \citep{2010AJ....140.1995L}.  
But there is significant
opportunity to identify RTs through analysis of archival 
radio data.  \citet{2007ApJ...666..346B} used nearly 1000 observations of a blank field
observed by the Very Large Array (VLA) over 20 years to identify a set
of RTs that have no apparent counterpart at radio or optical
wavelengths.  \citet{2010ApJ...711..517O} have suggested that these may be due to 
neutron stars.  \citet{2002ApJ...576..923L} and
\citet{2006ApJ...639..331G} conducted a search for RTs
at 1.4 GHz through a comparison of the VLA NVSS and FIRST survey catalogs 
that identified a radio supernova in the nearby galaxy NGC 4216.
\citet{2010arXiv1011.0003B} recently completed a search at 843 MHz of the Molongolo Observatory
Synthesis Telescope (MOST) data
archives that uncovered 15 RTs and a larger number of variable
sources.  Some of the sources discovered in the Molongolo search appear similar to
the RTs found by \citet{2007ApJ...666..346B} in that they have no faint radio or optical
counterpart.  \citet{2010AJ....140..157B} found a population of faint, variable radio sources
in the galactic plane, the majority of which are without multi-wavelength counterparts.

We describe here a search for radio transients in the field of the 
quasar 3C 286 with analysis of 1.4 GHz archival data from the VLA.  
One of the standard flux calibrators for the VLA, 3C 286 has been observed
thousands of times over the life time of the array.
This search
builds on the work of \citet{2007ApJ...666..346B}; however, observations in the 
vicinity of the bright (15 Jy) source 3C 286 place limits on 
the sensitivity that can be achieved.  The dynamic range of these
observations is limited by calibration errors and other systematic
effects rather than by statistical noise.  These calibration and
systematic errors can produce an apparent source in the image that
is well above the theoretical detection threshold.  
Accordingly, we require stronger evidence (such as appearance in 
consecutive epochs and comparisons with other cataloged sources
in the field) to demonstrate the existence
of an RT.  We present the data and its analysis in \S 2, a simulation demonstrating
the ability to identify sources with these methods in \S 3, our source detections
and transient identification efforts in \S 4, limits in transient surface density in \S 5,
and a summary in \S 6.

\section{Data and Analysis}

All of the data used were archived VLA observations. 
We limited our search to 1.4-GHz continuum-mode observations 
in C and D configurations.  Continuum mode has a bandwidth of 50 MHz 
in two separate intermediate frequency bands and in both 
circular polarizations.  In the longer baseline VLA configurations (A and B), 
bandwidth smearing reduces the area that can be imaged without significant
losses.  We selected all observations between April 1984 and May 2007.
After culling epochs with 
corrupt or too few data, there were 1852 observations to inspect.
The median duration of the epochs is 2 minutes; 0.2\% of epochs have a duration
shorter than 20 seconds and 0.7\% of epochs have a duration longer
than 20 minutes.  The total integration time included is 167 hours.

We used an automated flagging, calibrating and imaging pipeline 
within the AIPS package
originally developed for another archival field \citep{2007ApJ...666..346B}.  
We self-calibrated each epoch assuming a point source at the field center.
As demonstrated in \S 3, this method permits recovery of other sources
over a wide range of flux densities.
Each epoch was then imaged over a square area slightly larger than two times
the primary beam diameter.
The presence of a bright calibrator (3C 286) at the center of the field
permits us to perform amplitude and phase self-calibration on the data.

\begin{figure}
\subfigure{\epsfig{file=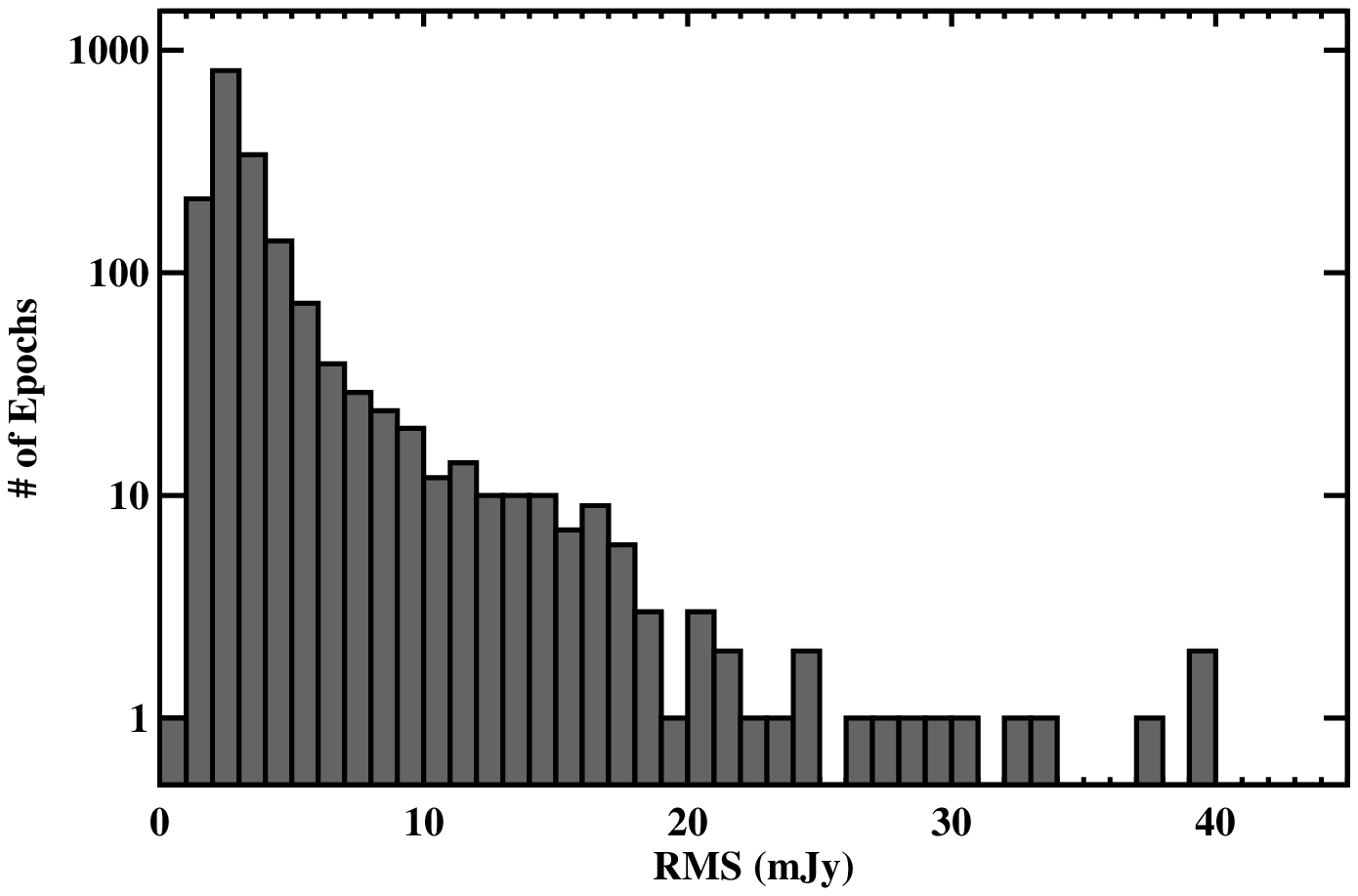,scale=0.5}}
\subfigure{\epsfig{file=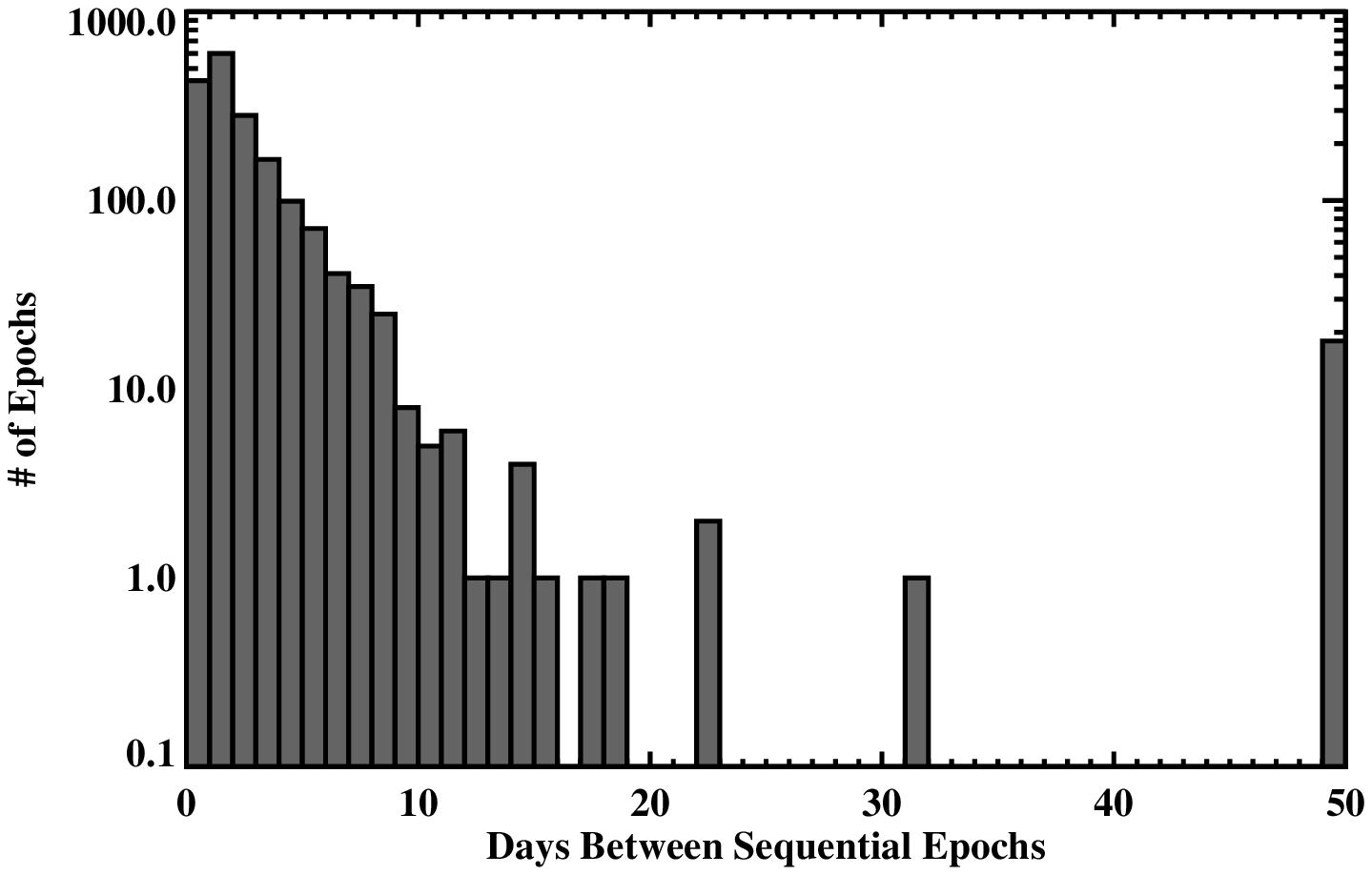,scale=0.5}}
\caption{Histograms of survey properties. {\it Left:} Histogram of epoch RMS noise levels. {\it Right:} Histogram of number of days between 
sequential epochs. Multiple observations on the same day are considered to have a zero day 
separation while all separations greater than 50 days are included in the rightmost bin for the sake 
of clarity. \label{fig:stats}}
\end{figure}

The median rms flux density of individual epoch images is 2 mJy, corresponding
to a dynamic range of $\sim 10^4$ (Fig.~\ref{fig:stats}).  We also plot the
histogram of separation between sequential epochs.  The characteristic
separation between epochs is $\sim 1$ day.  Typical resolution in the
C and D configurations is $\sim 15$ arcsec and 45 arcsec, respectively.

\begin{figure}
\epsfig{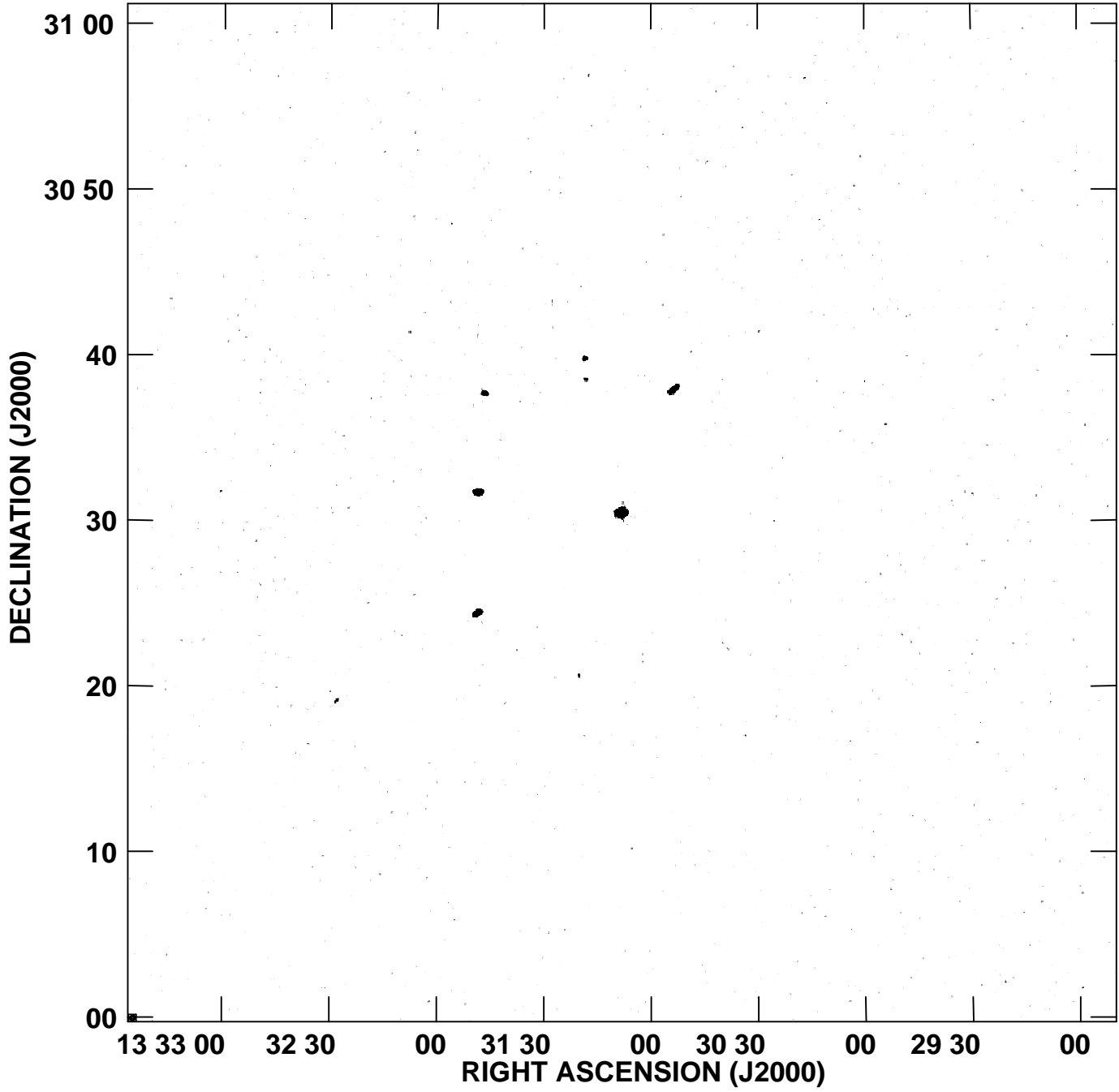}
\caption{Deep image of the 3C 286 field.  3C 286 is at the center of the field.  The 
gray scale goes from 2 to 4 mJy.
\label{fig:deep}}
\end{figure}

We created a deep image (Fig.~\ref{fig:deep})
by combining 1517 of the epochs with the 
MIRIAD software package. The rms of the deep image excluding bright sources
is 0.8 mJy and has a synthesized beam of $22 \times 16$ arcsec$^2$.  
The image is dynamic range limited and probably could be improved with a significant
self-calibration effort.  However, the image is very effective at identifying 
sources in the field that may show up in individual epoch images.
We identify 10 sources in the field including 3C 286 (Table~\ref{tab:steady}).
As discussed below, we tabulate the detections of each of these sources in
all of the epochs.
We report the mean flux density over all the epochs, the modulation (i.e., root mean square
variation) of that flux density, and the
number of epochs in which sources are detected ($N_{det}$).  For sources with $N_{det}=0$,
we report the flux density from the deep image as the mean flux.
The overdensity of strong sources in the vicinity of 3C 286 gives a 
useful benchmark against systematic errors.

\section{Simulation}

A concern in analysis of data of this kind is that self-calibration on the bright
point source in the field will distort or disappear a transient source that is not
included in the self-calibration model.  The persistence of the other sources in the
field demonstrates that self-calibration is robust against this problem for 
sources with fluxes between 30 and 200 mJy.  We performed a simple simulation to explore
the effect on sources with flux densities from 10 mJy to 25 Jy.  The simulation
creates fake data using the MIRIAD task UVGEN for the VLA D configuration of a
2-minute 1.4 GHz snapshot observation of 3C 286 at transit.  The model sky is
composed of the deep sources in Table~\ref{tab:steady} with an additional
source at a separation of 700 arcsec from the field center with flux $S_{fake}$.  Self-calibration is
performed with a simple model of a point source at the phase center.  The data
are then imaged, cleaned, and restored.  The flux density of the source at the phase
center is consistent with the 3C 286 flux density until $S_{fake} \sim S_{3C286}$.
The fake source is recovered consistently at the input flux density 
for $S_{fake} < 10$ Jy.
This simulation confirms that we can recover RTs in the
field over a wide range of flux densities.

\section{RT Identification}

The AIPS task SAD (Search And Destroy) was used to identify all
sources in the fields brighter than $5\sigma$ over the full area
imaged.
We identified a total of 30067 sources in all epochs.  We perform
several exclusions to identify any transient sources.  We reject
any sources that are in the outer 2.5\% of the image which may 
be affected by edge problems in the image.  We reject any sources
outside of the two times the primary beam radius.  We reject any sources
with a fitted size larger than 120 arcsec since we are only interested in
point-like sources.  We reject any sources within 0.08 deg of 3C 286.
The latter step removes a large number of sidelobes associated with
the bright point source.  We also remove any sources that are matched
(within a radius of 30\arcsec)
to the known steady sources in Table~\ref{tab:steady}; we tabulate the
number of $7\sigma$ detections for each of these sources, $N_{det}$.  Three
sources are detected in nearly all of the epochs; the faintest has
a flux density of 73 mJy.  Variations in the flux density indicate
systematic uncertainty of $\sim 5\%$.  None of the sources show strong
intrinsic variability.

After these cuts, we are left with 608 sources in the catalog.
186 of these sources are found to be repeating in different epochs
and 3 are found to be repeating in consecutive epochs.  
Examination of the images reveals that many of these
sources are associated with sidelobes of the synthesized beam.  
If we apply a higher threshold of $7\sigma$, then we reduce the number of 
single epoch candidates to 78.  Only 7 candidates are seen to repeat in any
epoch and none are repeating in consecutive epochs.  The repeating candidates
all cluster near the brightest of the steady sources (J133148+303148) and
are likely to be sidelobes.

Assuming purely Gaussian noise,
the expected number of false positive sources given a statistical threshold
$\sigma$
for all epochs is $N_{fp}=1/2  {\rm\ erfc}(\sigma/\sqrt{2})  N_{try} $,
where erfc is the complementary error function.
$N_{try}=N_{epoch} \Omega_f / \Omega_b$ is the product of the number
of epochs with the ratio of the search area to the synthesized beam area.
For this experiment, $N_{try} \approx 6\times 10^7$.  For $\sigma=5$,
we expect $N_{fp}\approx 20$.  For $\sigma=7$, $N_{fp} \approx 8 \times 10^{-5}$.  
Thus, the remaining $7\sigma$
candidates are either real RTs or they are systematic errors.  
Establishing the reality of any single epoch candidate is very difficult given the
variable systematic errors present in the data.  Thus, at the low sensitivity end, we primarily
make use of only the 
consecutive event counts to estimate transient sensitivity.  This has a characteristic
time scale of the epoch separation $\sim 1$ day.  At higher sensitivity, the 
characteristic RT time scale sampled is $\sim 1$ minute.

We note one unusual candidate source that showed up in a single epoch (13 July 1996).  The source
appeared as a bright (422 mJy) source with point-like structure.  The steady sources were
present in the image at the right flux densities.  This candidate
had the appearance of a convincing
RT.  However, examination of the visibility data during this 2-minute observation
indicated a glitch.  It appears that the online flagging system was effective in removing
most of the bad data but some bad data remained.  In those bad data, the phase center
of the image was shifted.  Imaging all of the data together led to point sources at the
position of 3C 286 and at the shifted phase center.  The result was an apparently convincing
RT candidate until we removed two 10-second integrations.  Note that this glitch differs 
from the end of record problem identified by \citet{2010ApJ...711..517O}.
After exclusion of this event and four other single epoch candidates in which imaging revealed poorly flagged and/or poorly calibrated data, there were no single epoch candidates brighter than 250 mJy.

\section{Transient Rate Estimates}

\begin{figure}
\epsfig{figure=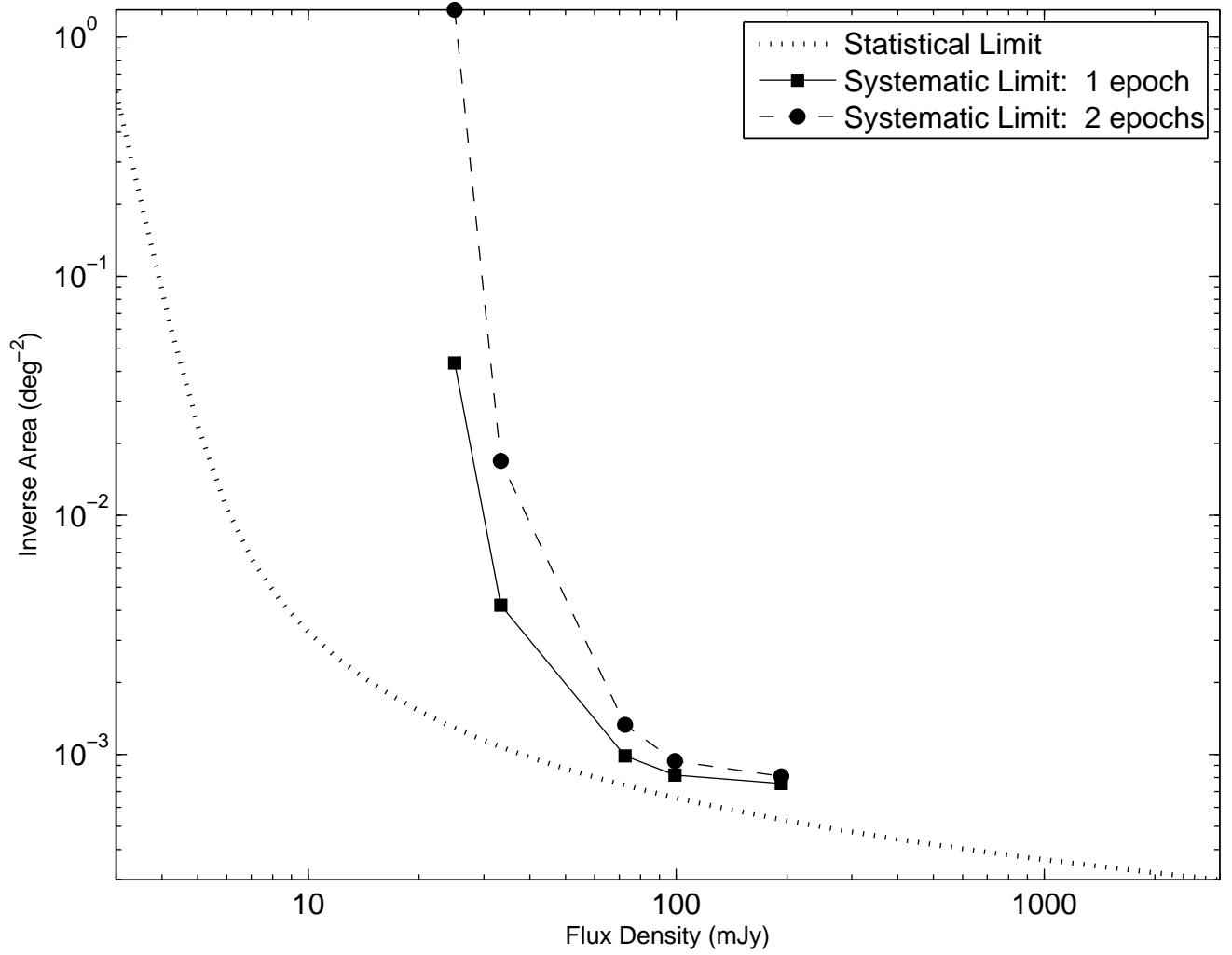}
\caption[]{Inverse area for statistical and systematic limits as a function of flux
density threshold.  See text for details of this calculation.
\label{fig:inverse}
}
\end{figure}

We calculate the transient surface density, $\Sigma$,
which is equivalent to the two-epoch transient rate, $R$, given in earlier papers  
\citep{2007ApJ...666..346B,2010ApJ...719...45C,2010ApJ...725.1792B}.
The key step is to estimate the
area per epoch as a function of flux density threshold.  We can estimate this based on 
the image rms and the imaged area with a sensitivity above the threshold; this is 
the statistical limit given in Fig.~\ref{fig:inverse}.  We plot here the inverse of
the area, which is proportional to the surface density.  This statistical method, however,
ignores the effects of systematic errors caused by 3C 286 in the image.  We use an alternative
method of estimating the area available by counting the number of epochs in which the bright
steady sources in the image are detected (Table~\ref{tab:steady}).  Under the assumption
that this threshold is much higher than the statistical noise threshold (which holds in
most cases), then the total area
is the number of detections times the field of view imaged; this is labeled as 
systematic limit for 1 epoch.  We can also use
the requirement of consecutive detections of the steady sources to estimate the area for repeating
sources; this is the systematic limit for 2 epoch detections.  The inverse area 
rises steeply at flux densities below 70 mJy.  For flux
densities greater than 70 mJy, we find good agreement between the statistical and systematic
approaches and little difference between the one and two epoch areas.
The discrepancy between the statistical and systematic limits at low flux densities is the
penalty that we pay for using data with a dynamic range limit.  

\begin{figure}
\epsfig{figure=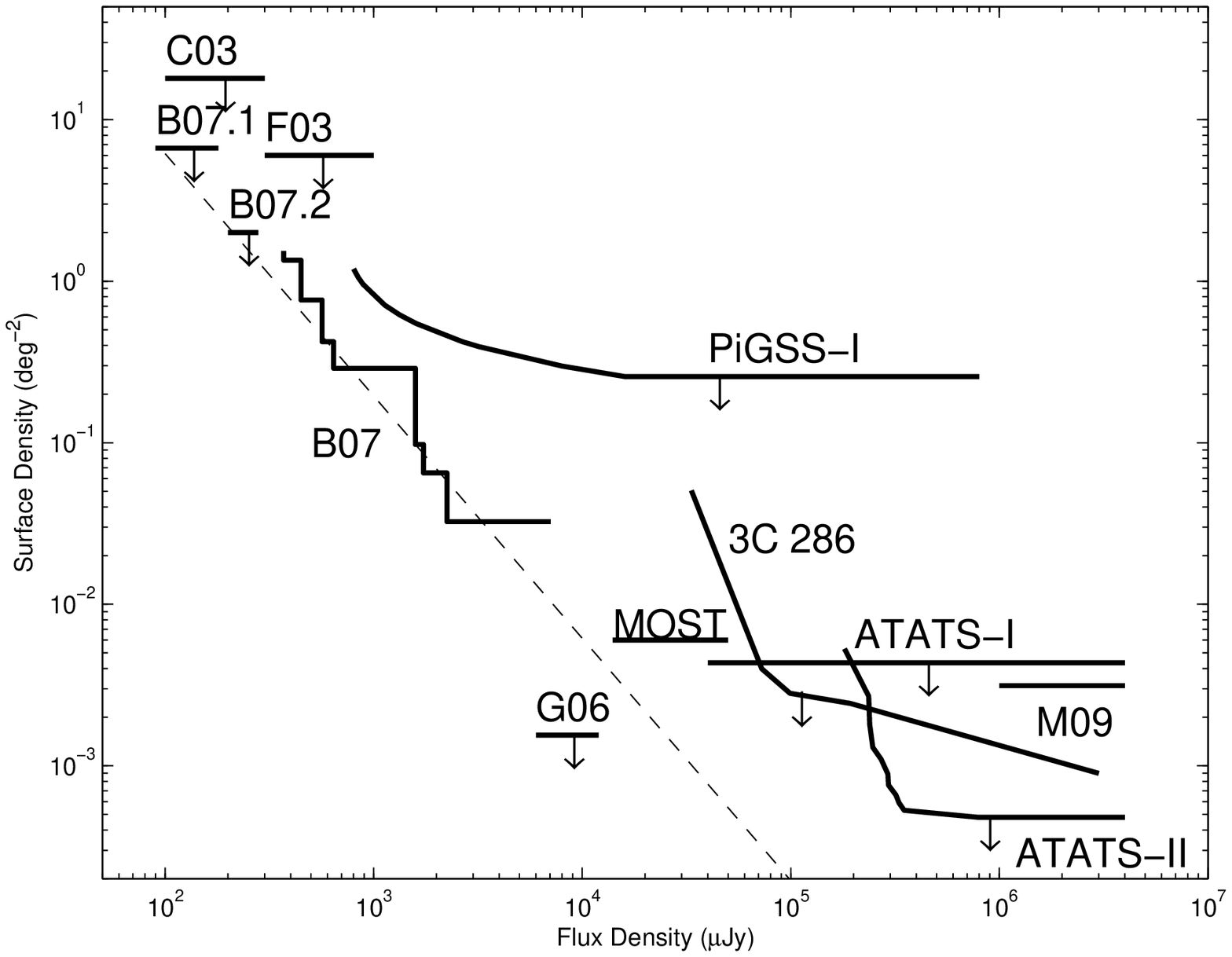,width=\textwidth}
\caption[]{
Transient surface density
from this and other surveys as a function
of flux density.  Result from this survey is labeled 3C 286.
Curves and lines indicate detected values and upper limits from 
a deep VLA search \citep[B07,B07.1,B07.2;][]{2007ApJ...666..346B},
the comparison of the 1.4~GHz NVSS and FIRST surveys
\citep[G06;][]{2006ApJ...639..331G}, from additional
VLA searches \citep[C03 and F03;][]{2003ApJ...590..192C,2003AJ....125.2299F},
 from the first and second ATATS papers 
\citep[ATATS-I and ATATS-II;][]{2010ApJ...719...45C,ATATSII},
from the first data release of PiGSS \citep[PiGSS-I;][]{2010ApJ...725.1792B}.
from the \citet[M09;][]{2009AJ....138..787M} survey, and from
the MOST search \citep{2010arXiv1011.0003B}.  The dashed line is proportional to $S^{-1.5}$ and
is normalized to B07 estimates.  Lines with arrows indicate $1\sigma$ upper limits; otherwise
the results are indicative of detected transients (B07, MOST, and M09).
\label{fig:rate}
}
\end{figure}

We estimate the $1\sigma$ upper limit on $\Sigma$ and plot it against other measurments and limits 
(Figure~\ref{fig:rate}).
We use the systematic two-epoch limit for flux densities $< 200$ mJy and then interpolate
to the statistical limit for very large flux densities.  The surface density above 200 mJy applies
to transients with timescales $>1$ minute.  We find $\Sigma < 3 \times 10^{-3}$ deg$^{-2}$
at 70 mJy (for $\tau \sim 1$ day) and $\Sigma \approx 9 \times 10^{-4}$ deg$^{-2}$ at 3 Jy
(for $\tau \sim 1$ min).  

These limits are an improvement over limits on very bright RTs relative to the ATATS-I
and MOST surveys 
\citep{2010ApJ...719...45C,2010arXiv1011.0003B}
and are comparable to the limits from the NVSS-FIRST comparison \citep{2006ApJ...639..331G}.  
The ATATS-II results are a factor of $\sim 3$ more sensitive than the 3C 286 results
at a flux density of 300 mJy \citep{ATATSII}; however, the 3C 286 results are a factor of 4 times
more sensitive in flux density.  Our limits from 3C 286 observations
are more sensitive than the published  limits from \citet[][M09]{2009AJ....138..787M}, however, there is considerable uncertainty about
what is the best limit based on the M09 data.  Previous limits from the same group
\citep{2008NewA...13..519K} also disagree with M09.  We plot the stated M09 surface density 
from two years of surveying  of 
$8.7\times 10^{-7}\ $ arcmin$^{-2} = 3\times 10^{-3}$ deg$^{-2}$.
Our independent calculation based on
data presented in the paper of 3 transients detected in 50 days of observing with an
instantaneous field of view of 500 deg$^2$ suggests a surface density of $10^{-4}$ deg$^{-2}$; 
using 9 transients from 2 years of observing we estimate $\Sigma\sim 2 \times 10^{-5}$ deg$^{-2}$.
However, we cannot be certain that we know the duty cycle of this observing and so take the 
stated M09 limit of $3 \times 10^{-3}$ deg$^{-2}$ as the benchmark value.  For this value, we
would expect to find 3 1-Jy RTs, which we do not find.  

The
surveys compared in this plot cover an order of magnitude in frequency (0.84 to 8.4 GHz) 
and several orders of magnitude in time scale that are probed (1 minute to 1 year).
Thus, each survey takes a different slice of parameter space that may probe very different
physics and source populations.  Broadly, these observations are sensitive to synchrotron
phenomena, such as the explosive ejecta of radio supernova and gamma-ray burst afterglows, 
as well as X-ray binaries and active galactic nuclei.  Models for most of these source
populations indicate surface densities that are to the left and below the 
dashed line in the plot \citep[e.g.,][]{2008MNRAS.390..675R}.

\section{Summary}

We have presented an analysis of 1852 epochs of VLA observations spanning 23 years
of the bright calibrator 3C 286 at 1.4 GHz.  This data set provides an important 
search for radio transients brighter than 70 mJy.  We do not find any transients,
in contradiction with optimistic estimates of transient surface density
from M09 but consistent with
limits from other surveys.  Differences between surveys may be a function of 
observing frequency, regions of sky covered, and systematic problems in recovering
transients.

The results are limited significantly by dynamic range of the imaging.  If systematic
errors had not contributed to the imaging, our surface density limit would apply 
at flux densities that
are an order of magnitude lower.  Nevertheless, the results demonstrates that searches
around bright calibrators can provide unique information.  Future searches
of the VLA archives can improve on these results through the use of the larger number of
observations of other standard calibrators such as 3C 48 and of observations at other 
frequencies.  
The examination of fainter calibrators is an important way to get closer
to the statistical noise limits under the assumption of a fixed
dynamic range limit.  Finally,
more complete models used in self-calibration may permit higher dynamic range imaging.
If 1\% of the data from the VLA archives consist of calibrators suitable
for transient searching, we will have 2000 hours of usable data from the past 25 years,
corresponding to an order of magnitude increase in sensitivity to radio transients.

\acknowledgements{The National Radio Astronomy Observatory is a facility of the National Science Foundation operated under cooperative agreement by Associated Universities, Inc.}


\begin{deluxetable}{rrrrr}
\tablecaption{Persistent Sources \label{tab:steady}}
\tablehead{ \colhead{RA} & \colhead{Dec.} & \colhead{Mean Flux}
& \colhead{Modulation} & \colhead{$N_{det}$} \\
                              \colhead{ (J2000) } & \colhead{ (J2000) } & \colhead{(mJy)} & \colhead{(mJy)} & }
\startdata
13:32:04.7 & 30:21:15.4 &     4 & \dots &    0 \\ 
13:31:48.7 & 30:24:29.7 &    99 &    13 & 1587 \\ 
13:31:48.4 & 30:31:47.5 &   193 &    23 & 1720 \\ 
13:31:46.6 & 30:37:46.6 &    33 &     6 &  309 \\ 
13:31:20.3 & 30:20:41.9 &    15 & \dots &    0 \\ 
13:31:18.5 & 30:39:52.2 &    25 &     7 &   30 \\ 
13:31:18.3 & 30:38:36.0 &     8 & \dots &    0 \\ 
13:31:08.3 & 30:30:32.9 & 15488 & \dots &    \dots \\ 
13:30:53.6 & 30:38:01.0 &    73 &    16 & 1320 \\ 
13:30:50.1 & 30:27:30.8 &     3 & \dots &    0 \\ 
\enddata
\end{deluxetable}

\end{document}